\documentclass[aps,reprint,longbibliography,twocolumn]{revtex4-1}

\usepackage{bm}

\usepackage{amsmath,amssymb}
\usepackage{graphicx,color,dsfont}
\usepackage{etoolbox,mathtools}
\usepackage[titletoc,toc]{appendix}

\usepackage{lineno} 

\DeclareGraphicsExtensions{.pdf,.png,.jpg}
\begin{document}

\title{Exploring Temporal Networks with Greedy Walks}

\author{Jari Saram\"aki}\email{jari.saramaki@aalto.fi}
\affiliation{Department of Computer Science, Aalto University School of Science, P.O. Box 15400, FI-00076 AALTO, Finland}\author{Petter Holme}
\affiliation{Department of Energy Science, Sungkyunkwan University, Suwon 440-746, Korea}
\date{\today}

\begin{abstract}{
Temporal networks come with a wide variety of heterogeneities, from burstiness of event sequences to correlations between timings
of node and link activations. In this paper, we set to explore the latter by using \emph{greedy walks} as probes of temporal network structure. Given a temporal network (a sequence of contacts), greedy walks proceed from node to node by always following the first available contact. Because of this, their structure is particularly sensitive to temporal-topological patterns involving repeated contacts between sets of nodes. This becomes evident in their small coverage per step  as compared to a temporal reference model -- in empirical temporal networks, greedy walks often get stuck within small sets of nodes because of correlated contact patterns. While this may also happen in static networks that have pronounced community structure, the use of the temporal reference model takes the underlying static network structure out of the equation and indicates that there is a purely temporal reason for the observations. 
Further analysis of the structure of greedy walks indicates that burst trains, sequences of repeated contacts between node pairs, are the dominant factor. However, there are larger patterns too, as shown with non-back\-track\-ing greedy walks.  We proceed further to study the entropy rates of greedy walks, and show that the sequences of visited nodes are more structured and predictable in original data as compared to temporally uncorrelated references. Taken together, these results indicate a richness of correlated temporal-topological patterns in temporal networks.
}
\end{abstract}

\maketitle


\section{Introduction}

When it comes to complex networks, temporal networks truly deserve to be called complex, because of their wide range of heterogeneities~\cite{saramaki_holme_2012}. While they inherit common structural heterogeneities of static networks such as clustering and communities, they also exhibit purely temporal heterogeneities, \emph{e.g.}\  burstiness of contact sequences~\cite{Barabasi2005,Vazquez2007,Iribarren2009,Karsai2011,Karsai2012a}. There are also structures that could be categorised as temporal-topological, such as temporal subgraphs and motifs~\cite{Kovanen2011,Kovanen2013} that consist of rapid sequences of contacts within small sets of nodes. Temporal motifs can be viewed as a subset of an even larger class of higher-order temporal structures, where the contacts of a sequence are both temporally and structurally correlated and of a non-Markovian nature~\cite{Scholtes2014}: future contacts depend on when and where past contacts happened. This class also includes triggered events where events in the neighbourhood of a node are seen to frequently follow one another within a short period of time (see, e.g., \cite{Backlund2014}), the phenomenon of betweenness preference~\cite{betweenness_preference}, where events typically follow certain local pathways, and the frequently occurring burst trains of events (``ping-pong patterns'') between pairs of nodes in communication networks~\cite{KarsaiCorrelated,zhao_etal}. 

\begin{figure}[!ht]
\begin{center}
\includegraphics[width=0.8\linewidth]{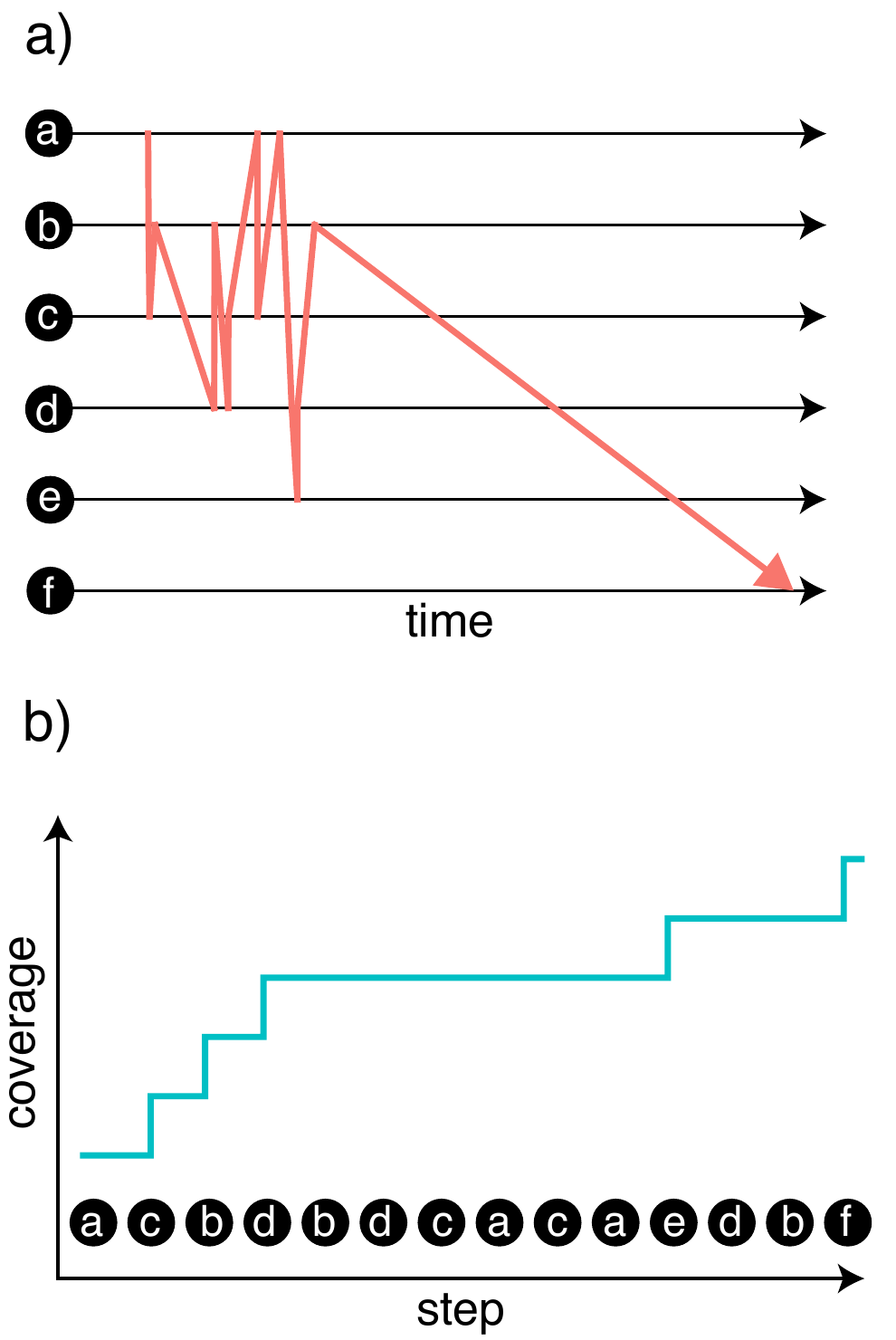}
\end{center}
\caption{The anatomy of a typical greedy walk, as illustrated with the first steps of a greedy walk in one of our data sets. a) The trajectory of the walk as a function of time. Each horizontal line indicates the timeline of a node, and the walk follows the pink path. The trapping effect typical for our data sets is clearly visible, as the walker repeatedly visits a small number of nodes. b) Coverage (the number of distinct nodes visited) as function of steps taken for the same walk.}
\label{fig:example}
\end{figure}

In this paper, we set out to investigate temporal-topol\-ogical structures spanned by consecutive events between nodes. We introduce the concept of temporal \emph{greedy walks}, walks that are purely determined by the sequence of events in the temporal network that acts as a substrate. We then use such walks as probes of temporal network structure. Temporal greedy walks have no counterpart in static networks. A greedy walker on a temporal network always follows the first event out of its current node. Therefore, the path it takes is particularly sensitive to temporal correlations involving adjacent links, in particular the above-mentioned non-Markovian contact sequences containing repeated contacts with small groups of nodes, from burst trains to temporal motifs. Because such temp\-oral-topol\-ogical patterns then trap the walkers within these node groups,  analysis of the structure of the paths taken by greedy walkers should reveal traces of these patterns (see Fig.~\ref{fig:example} a). In particular, comparing the properties of temporal greedy walks to their properties on reference networks, where such patterns have been removed with the help of time-stamp shuffling, allows estimating how dominant these patterns are in the temporal network structure. One advantage of this approach compared to e.g.~the temporal motifs approach~\cite{Kovanen2011,Kovanen2013} is that one does not need to specify patterns of interest beforehand; \emph{any} sequence of contacts that immediately follow one another counts.

Greedy walks are a limiting case of random walks~\cite{perra2012random,starnini2012,RochaNJP2014,delvenne2015}.
For pre-determined temporal network structure, such as empirical contact lists with time stamps, temporal greedy walks are entirely deterministic once the initial conditions (first node, time) have been set, as long as nodes only participate in a single event at a time, which 
is mainly the case with our empirical data. Note that in some temporal-network models of random walks~\cite{speidel_lambiotte,hoffmann_rw,egu}, the walks themselves are in fact temporally greedy, and randomness only comes from a stochastic model of the underlying temporal network. Further, in studies of random walks on temporal networks the focus has mainly been on issues such as effects of burstiness on mean first passage and relaxation times~\cite{starnini2012,delvenne2015}, models that generate temporal networks~\cite{barrat_itineraries}, and identification of timescales~\cite{delvenne_timescales,baronchelli_resolution}.
We believe that our work is the first to apply temporal greedy walks to analysis of empirical data on temporal networks. We take ``real time'' out of the equation (cf.\ Ref.~\cite{albano_ex_intrinsic}) and focus on the \emph{structure} of greedy walks, i.e.\ their order of visited nodes, step by step. We employ the commonly-used time-shuffled reference model in the same spirit, as a reference model that yields walks with temporally random event sequences that are not affected by timing correlations. 

\begin{figure*}[!ht]
\begin{center}
\includegraphics[width=0.75\linewidth]{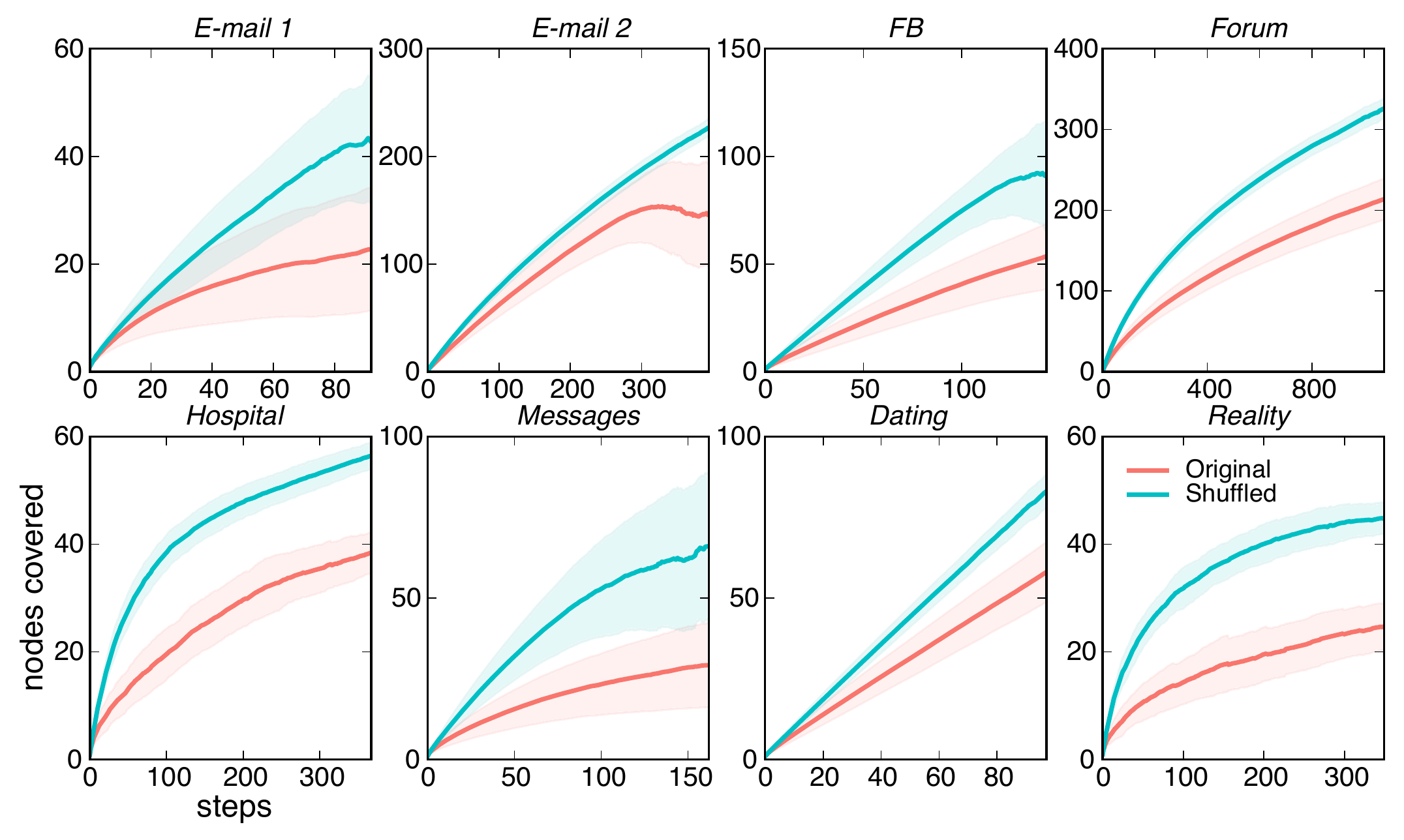}
\end{center}
\caption{The average number of unique nodes covered as a function of the number of steps taken for greedy walks in the eight empirical temporal networks. Red lines denote average coverage in original data, whereas blue lines are for time-shuffled reference networks. Shaded areas indicate standard deviations. Note that event timestamps have only been used to determine the order of events that the greedy walks follow, and here we do not consider the times taken between consecutive steps. Therefore, the explanation for the smaller coverage in original data is that the sequences contain correlated temporal patterns and chains of repeating events within small sets of nodes.}
\label{fig:coverage}
\end{figure*}

In this paper, we first investigate the coverage of greedy walks in empirical temporal networks (see Figure~\ref{fig:example} b). As the coverage of a walk measures the number of unique nodes visited, measuring its growth as a function of the number of steps taken is a good way of revealing the existence of ``traps'', where greedy walkers remain within a small set of nodes for prolonged times, because of burst trains, temporal motifs, and the like. The results of this analysis point out that there is an abundance of burst trains between pairs of nodes that dominate greedy walks. This is confirmed by the very high fraction of back\-track\-ing steps in the walks as compared to the reference model. Because of this, we next turn to non-back\-track\-ing walks, where the greedy walkers are not allowed to directly trace their last step back, and show that there are correlated temporal patterns beyond the burst trains. Finally, we apply an information-theoretic measure to the greedy walks, and show that for both ordinary and non-back\-tracking walks, the entropy rates of the walks are typically lower than in the reference networks.

\section{Data sets and simulations}

We study simulated greedy walks on eight different temporal network data sets that contain time-stamped events between nodes. Six of the data sets are electronic records of e-mail communication  (\emph{E-mail 1}, \emph{E-mail 2}) to Internet communities (\emph{FB}, \emph{Forum}, \emph{Messages}, \emph{Dating}) and two represent physical proximity (\emph{Reality}, \emph{Hospital}). For details, see Table~\ref{table:table1}.

\begin{table}

\begin{tabular}{l|rrrrrr}
Name  & $N$ & $E$ & $T$ & $\Delta t$  \\
\hline
{\em E-mail 1}~\cite{ebel2002scale} & 56,576 & 431,138 & 112 d & 1 s  \\
{\em E-mail 2}~\cite{eckmann2004entropy} & 3,186 & 308,726 & 82 d & 1 s  \\
{\em FB}*~\cite{viswanath2009evolution} & 31,359 & 566,305 & 15,000 h & 1 s  \\
{\em Forum}*~\cite{karimi2014structural} & 6,625  & 1,359,075 & 2,400 d & 1 s  \\
{\em Hospital}~\cite{vanhems2013estimating} & 75 & 32,424 & 4 d & 20 s  \\
{\em Messages}*~\cite{karimi2014structural} & 22,695 & 280,717 & 3 d & 1 s  \\
{\em Dating}*~\cite{holme2004structure} & 17,009 & 185,578 & 250 d & 1 s  \\
{\em Reality}~\cite{Eagle08092009,betweenness_preference} & 64  & 13,131 & 8.6 h & 5 m
\end{tabular}

\caption{Properties of the temporal network data sets: $N$ -- the number of links, $E$ -- the number of events, $T$ -- time interval covered by the data, and $\Delta t$ -- time resolution of the data. Transient periods are removed from networks marked with an asterisk.}
\label{table:table1}
\end{table}
 
We have performed exhaustive simulations of greedy walks beginning at every node at time $t=0$, continued until the end of data, for the eight different empirical temporal networks detailed above. In each run, the greedy walker always follows the first available event out of its current node; if there are multiple simultaneous events (this happens in the physical proximity data sets), the walker randomly picks one (and therefore becomes a random walker for this particular step). The whole node sequence that the greedy walker follows is then recorded. Computationally, this is done by keeping track of the current location of the walker while looping through the time-ordered set of contacts of the temporal network, and moving the walker to the next available node once a corresponding contact (or set of contacts) is encountered. This makes computing one greedy walk scale as $O(E)$, where $E$ is the number of contact events. Note that our networks are small enough to allow exhaustive simulations even on a desktop computer; for larger data sets, randomly sampling starting nodes should be sufficient. In any case, the problem is embarrassingly parallel, and exhaustive computation for greedy walks on larger networks should be possible in a reasonable amount of time with a computing cluster. 

For reference, similar simulations have been performed using time-shuffled reference networks where the time stamps of all events are randomly exchanged. This procedure retains the original number of events between all pairs of nodes but removes all temporal correlations between events on adjacent links. In the exhaustive simulations that cover all nodes as starting points, the networks have been re-shuffled for every 500 nodes in order to save computation time; further reshuffling would not qualitatively change the results. For the smallest Hospital and Reality networks, time-shuffling was repeated for each greedy walk.

\section{Results}

\subsection{Coverage and burst trains}

We begin our analysis by investigating the coverage of greedy walks as a function of the number of steps, in all empirical networks. 
For Fig.~\ref{fig:coverage}, we have first counted the number of unique visited nodes as a function of the number of steps taken for each greedy walk, and then computed the average and standard deviation of this quantity for all numbers of steps. As the lengths of walks measured in steps show large variation, for some excessively long walks, we perform the measurement only up to the number of steps taken by at least 50 different walks in order to avoid a lack of statistics. Figure~\ref{fig:coverage} displays this node coverage as a function of steps taken for all data sets (red lines), together results for  the time-shuffled reference model (blue lines).

\begin{figure}[!ht]
{\begin{center}
\includegraphics[width=0.93\linewidth]{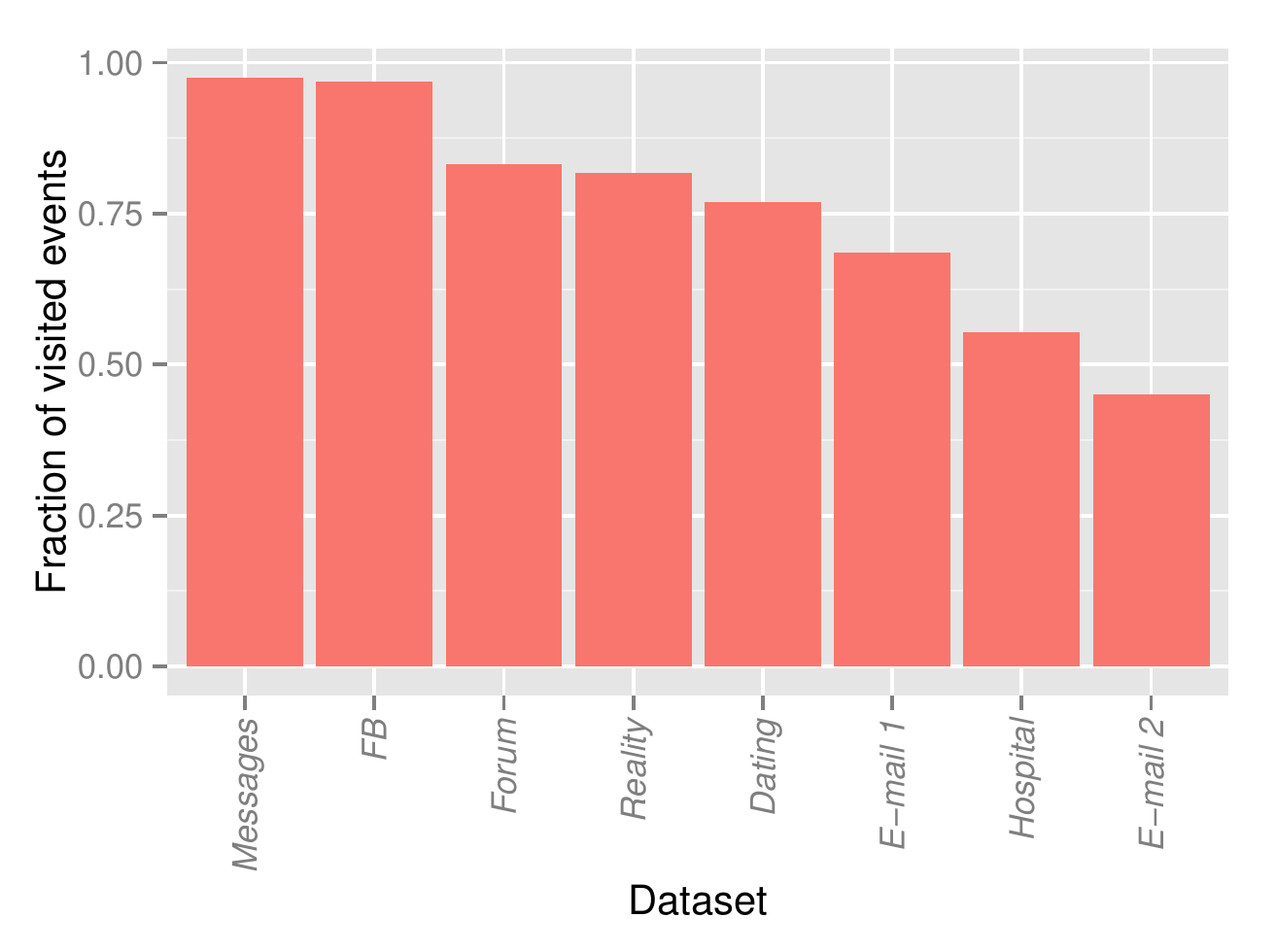}
\caption{The fraction of events through which at least one greedy walk passes, for each data set.}\label{eventcovers_1}
\end{center}}
\end{figure}

From Fig.~\ref{fig:coverage} it is clear that greedy walks on top of empirical event sequences, on average, cover fewer nodes per step than in the uncorrelated reference models\footnote{We have repeated the same analysis such that the greedy walks begin later, after one third of a data set's events have passed. This does not change our results.}. This means that the same nodes are visited more often in the empirical data compared to the time-shuffled reference sequences. Note that this difference between walks on the original networks and reference sequences only comes from temporal aspects like the non-Markovian nature of the original sequences -- because the underlying static network is the same for both original and reference sequences, topological features that may trap walkers (e.g.\ communities) are equally present in both sequences. Further, the observation cannot be directly attributed to the presence of burstiness: because we count steps, not how long it takes to take them, the observed slow growth of coverage does \emph{not} result from the slowing-down effects of burstiness that arise from long waiting times (as in e.g.~\cite{Karsai2011}).
However,
burst trains between node pairs and ``ping-pong patterns''~\cite{KarsaiCorrelated}, that is, burstiness that resides on links, do play a role in shaping the paths of greedy walks, as we will see below.

\begin{figure}[!ht]
{\begin{center}
\includegraphics[width=0.95\linewidth]{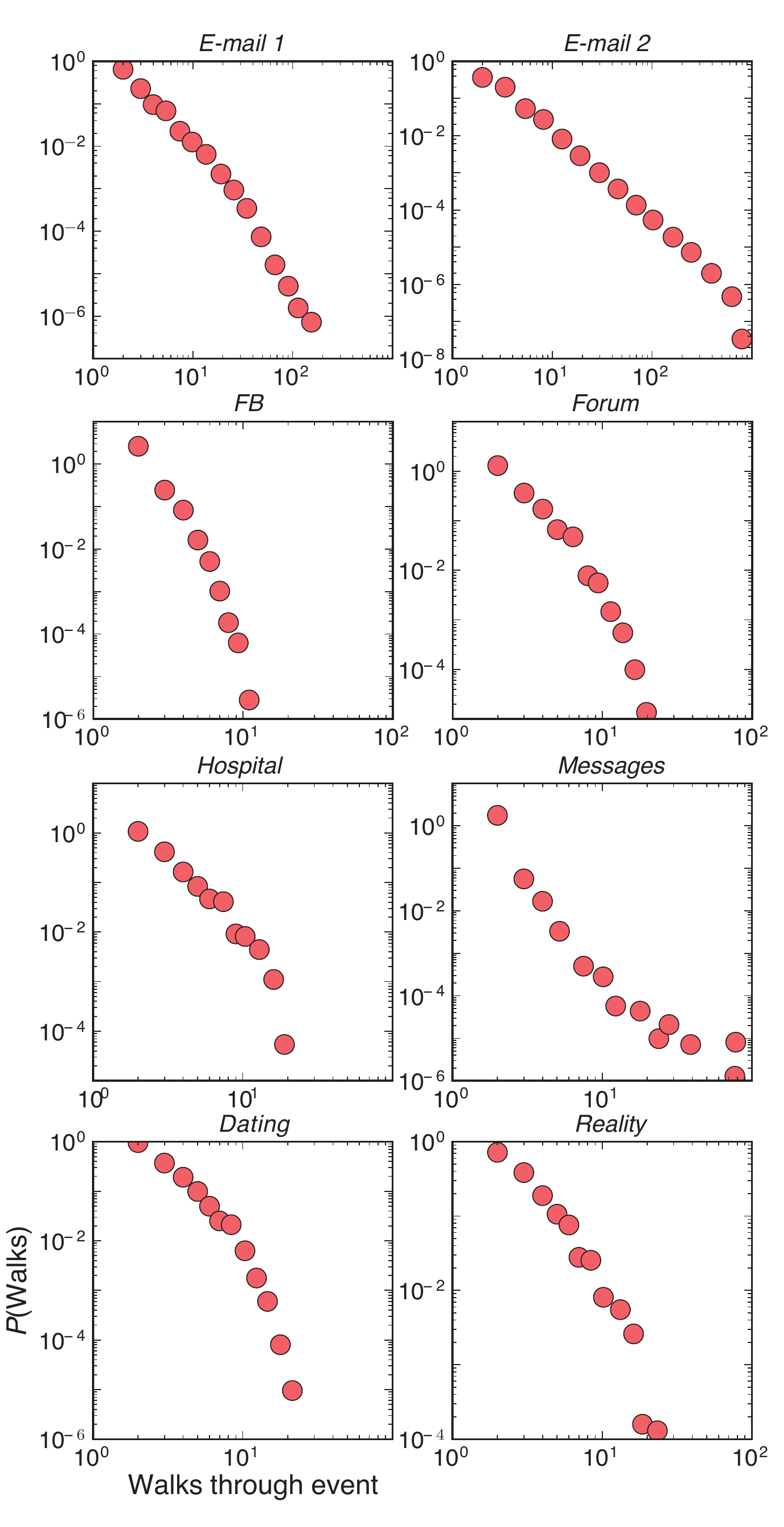}
\caption{The probability density function for the number of greedy walks that pass through each event, for each data set.}\label{eventcovers_2}
\end{center}}
\end{figure}

Because of the effects of trapping and broad event frequency distributions on links, not all events are touched by greedy walks. For our exhaustive simulations, the proportion of events that have carried greedy walkers from one node to the next range from 45$\%$ in \emph{Email-1} to 98$\%$ in \emph{Messages}, see Figure~\ref{eventcovers_1}. Moreover, counting the  number of greedy walks passing through each event, one sees that the numbers of walkers per event are clearly broadly distributed (Fig.~\ref{eventcovers_2}). This means that only a few walks pass through most events, and thus not all walks get trapped to the same paths. However, there are some (rare) central events, and consequently central paths, through which many greedy walks pass. Note that after passing through the same event, all greedy walks follow the same path to either the end of the walk, or until there are multiple simultaneous exits from a node. 

The low node coverage of greedy walks discussed above can in principle result from any temporal-topological correlations that limit the number of nodes visited by walkers, from burst trains on links to larger patterns of repeated consequent contacts between nodes that cause the walks to fold back on already visited nodes. In order to quantify the role of the first (burst trains), we compute the total fraction of \emph{back\-track\-ing} steps, where the walker directly returns to the node from which it arrived (e.g.\ ABA), for all data sets. These fractions are shown in Fig.~\ref{fig:bt_steps}a) for both the original data and the time-shuffled reference sequences. Fractions of back\-track\-ing steps range from 29$\%$ to 67$\%$, while they are much lower in the reference sequences\footnote{Note that the highest fractions occur for the physical proximity networks. This is natural -- physical proximity is measured by sensors at constant short intervals, and hence sequences such as ABABABAB are common. They only mean that A and B have been in proximity for a prolonged period of time.}. Therefore, it is evident that the back-and-forth ping-pong patterns of burst trains that trap walkers play a major role in the low coverage of greedy walks. Furthermore, because of their surprisingly high abundance, with a high likelihood, burst trains can be expected to play an important role in other types of dynamical processes that unfold on temporal networks as well.

\subsection{Non-backtracking greedy walks}

\begin{figure}[!t]
\begin{center}
\includegraphics[width=0.85\linewidth]{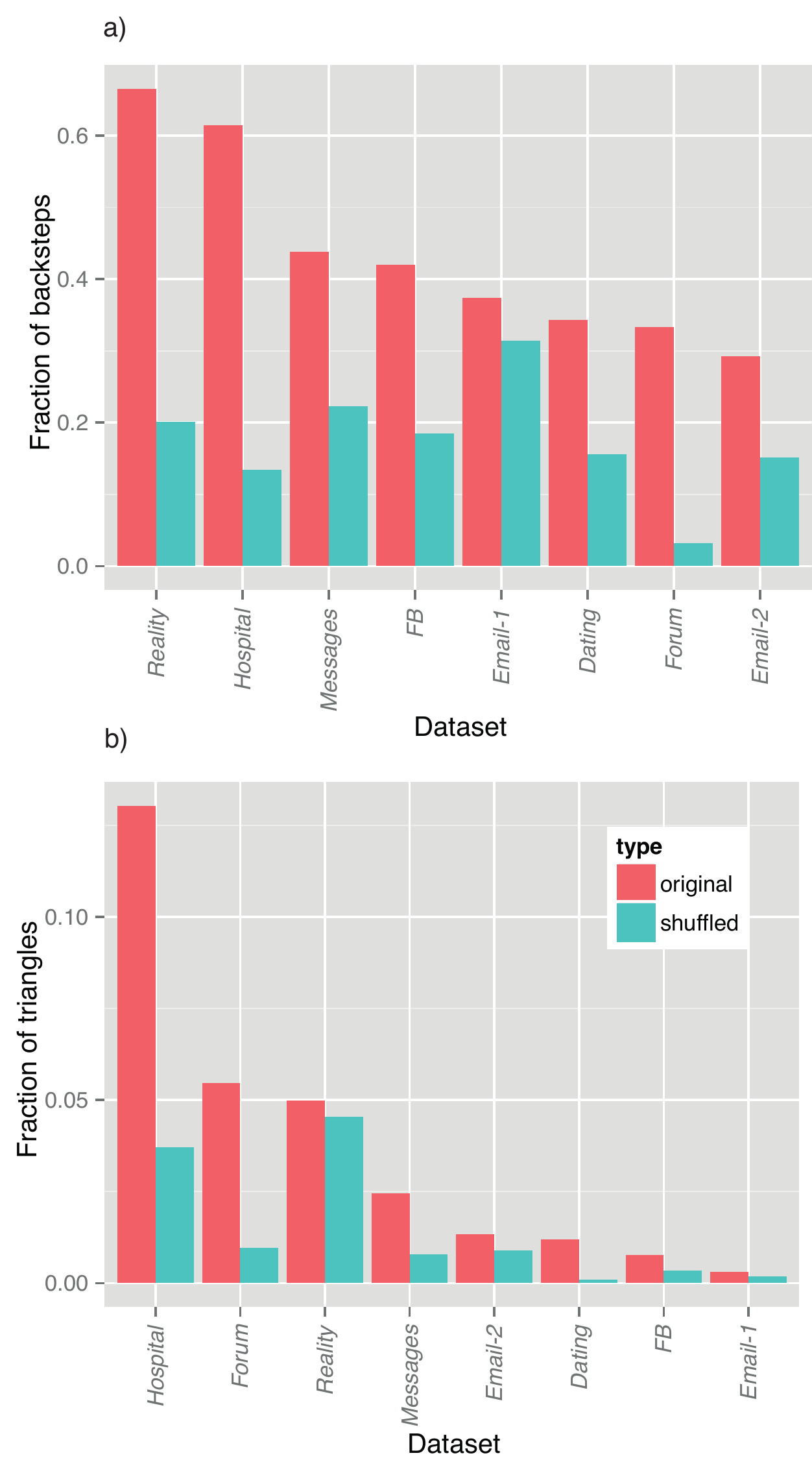}
\end{center}
\caption{a) The fraction of back\-track\-ing steps where the greedy walker immediately returns to the node it arrived from, in original and time-shuffled data sets. b) The fraction of triangle-closing steps (greedy walker returns to the node where it was two steps ago) in non-back\-track\-ing greedy walks, in original and time-shuffled data sets.}
\label{fig:bt_steps}
\end{figure}

Because burst trains in the shape of repeated contacts between node pairs are clearly a dominant factor in determining the coverage of greedy walks, we next deliberately disallow such patterns in order to understand the importance of larger temporal-topological structures. To this end, we simulate greedy walks that have one additional rule -- walkers always follow the next event out of a node that \emph{does not lead back to the previous node}. This means that these \emph{non-back\-track\-ing} greedy walks are not allowed to follow burst trains between two nodes; however, they may become trapped by any larger-scale patterns, from triangles (ABCA) to larger temporal motifs.

\begin{figure*}[!ht]
\begin{center}
\includegraphics[width=0.8\linewidth]{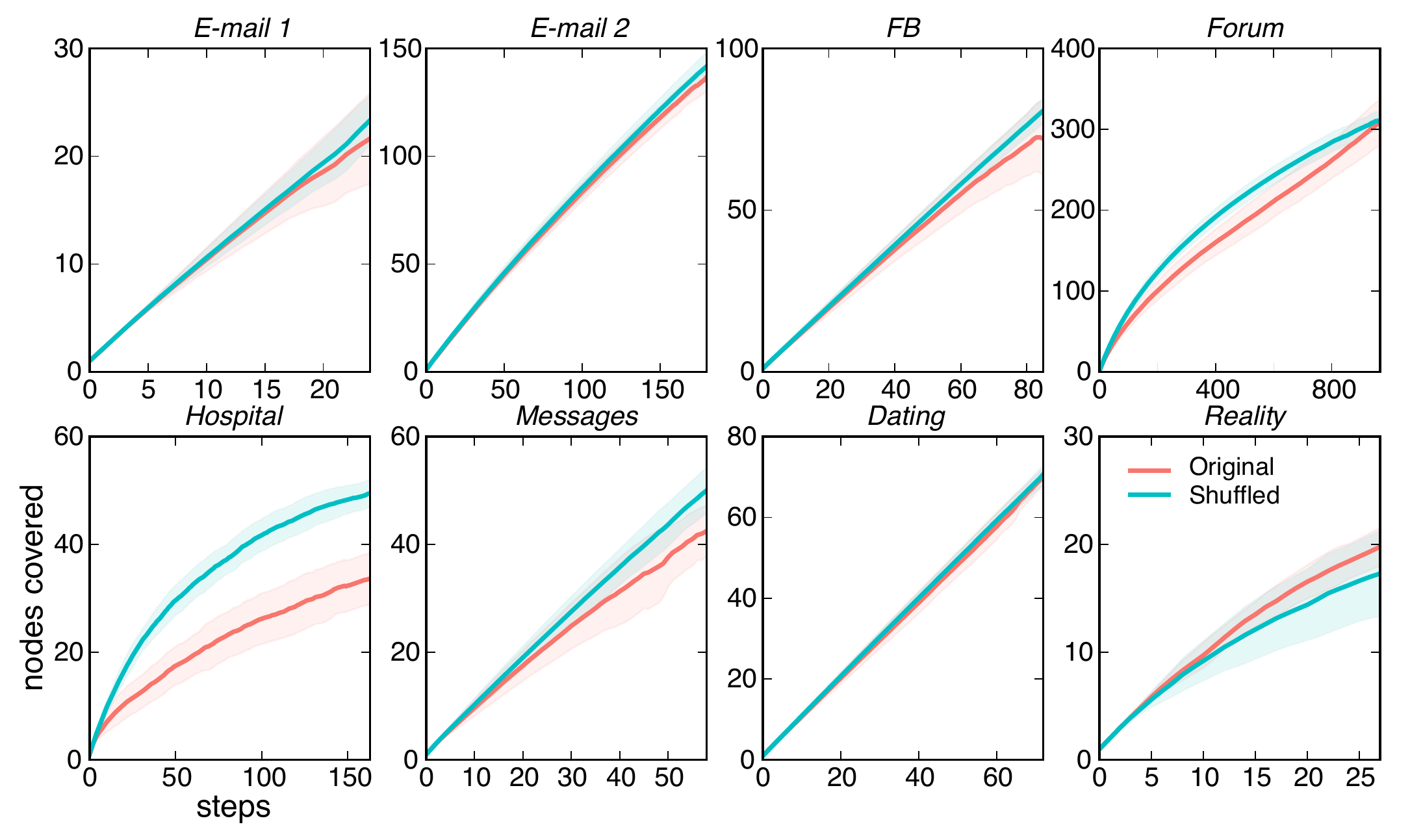}
\end{center}
\caption{The number of unique nodes covered as a function of the number of steps taken for non-back\-track\-ing greedy walks. Red lines denote coverage in original data, whereas blue lines are for time-shuffled reference networks.}
\label{fig:coverage_nonbt}
\end{figure*}

Figure~\ref{fig:coverage_nonbt} displays the average coverage of non-back\-track\-ing walkers as a function of the number of steps taken, for the original data and the reference sequences. Here, it is seen that for most data sets, there is still a difference, and the coverage grows more slowly in the original data (however, the difference is clearly smaller than for ordinary greedy walks). Thus, there are larger-scale topological-temporal structures (such as temporal triangles) in the original data that trap greedy walkers, albeit less frequently than the burst trains for ordinary walks. For \emph{Reality}, the difference is to the opposite direction: coverage in the original data grows faster than in the reference model. This is related to peculiarities of this small network. A visual analysis shows that it contains dense subnetworks with frequent events and lots of triangles; these dense subnetworks are active at different times, forcing non-backtracking greedy walkers to jump from one subnetwork to the next. In the time-shuffled reference networks, all subnetworks are active at all times and their abundance of triangles guarantees that walkers can repeatedly visit the same nodes (see also Fig.~\ref{fig:bt_steps}; the fraction of triangle-closing steps is very high in the shuffled reference for Reality). 

Similarly to the fraction of back\-track\-ing steps in ordinary greedy walks, we have computed the total fraction of triangle-closing steps in non-back\-track\-ing greedy walks for all data sets. These are steps that lead the walker to the node where it was two steps ago, e.g.\ the final step in ABCA is a triangle-closing step. The fraction of such steps is displayed in Fig.~\ref{fig:bt_steps}b) for all data sets. It is seen that although the fraction of triangle-closing steps is in general lower than that of back\-track\-ing steps, there is nevertheless a consistent difference between greedy walk structures in the original and shuffled data sets. This is indicative of the existence of larger temporal-topological structures from triangles to other motifs.

\subsection{Entropy rates of greedy walks}

We conclude by investigating the structure of greedy walks in more detail, and focus on quantifying the amount of repeated sequences in greedy walks. To this end, we
apply information-theoretic measures along the lines of Refs.~\cite{song2010,Takaguchi2011}. Specifically, we estimate the entropy rates $S$ of all greedy walks for both original and time-shuffled data following the approach of Song et al.~\cite{song2010}. The entropy rate of a sequence of symbols is defined as 
\begin{equation}
S = \lim_{n\rightarrow \infty}\frac{1}{n}\sum_{i=1}^nS(i),
\end{equation}
where $S(i)$ is the conditional entropy of the $i$'th step. For finite strings, one can estimate the entropy rate using the Lempel-Ziv estimator
\begin{equation}
\hat{S}=\left(\frac{1}{\ell}\sum_i\Lambda_i\right)^{-1}\ln \ell,
\end{equation}
where $\ell$ is the length of the sequence, and $\Lambda_i$ is the length of the shortest subsequence of visited nodes starting at step $i$ that does not appear previously
in the sequence. This estimator represents the 
asymptotic lower bound on the per-symbol description length when a realization of a stationary ergodic process is losslessly compressed~\cite{LempelZiv}. 
The estimator converges to $S$ when $n\rightarrow \infty$ if the source of the sequence is a stationary Markov chain of finite order~\cite{Shields1992,schurmann}; note that for non-Markovian sequences such as studied here convergence is not necessarily guaranteed. 
Because the formula assumes the sequence to be one-sided infinite (in order to compute the length of the shortest novel subsequence at step $i$), 
we have taken greedy walks of $L>20$ steps and computed the Lempel-Ziv estimator for $\ell=L/2$, i.e.\ the first half of the walk. 

%

\begin{figure*}[ht]
\begin{center}
\includegraphics[width=0.8\linewidth]{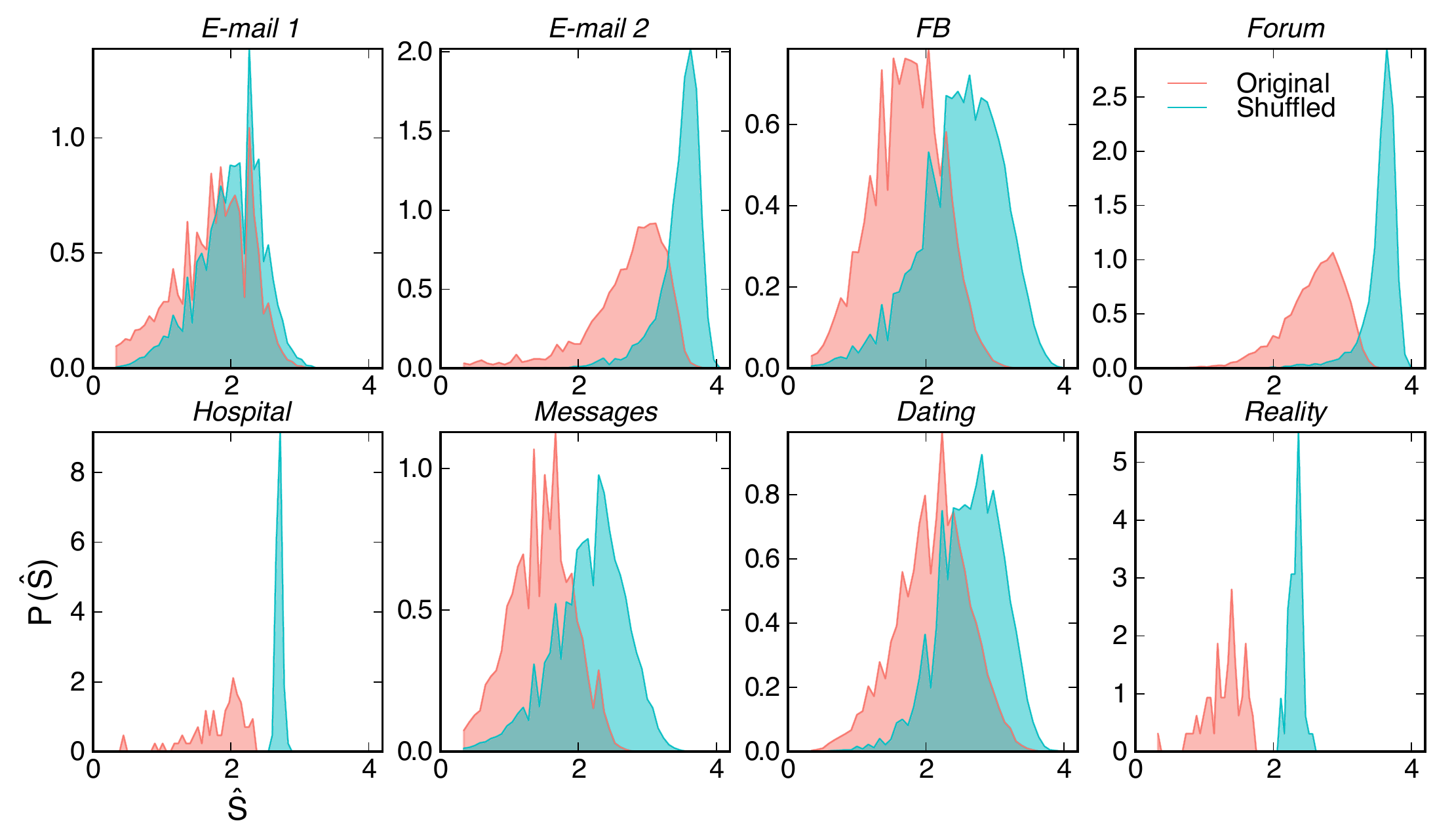}
\end{center}
\caption{Probability density functions for estimated entropy rates $\hat{S}$ of greedy walks for all data sets. The red lines denote PDFs for walks on top of original event sequences, whereas blue lines are again for time-shuffled reference data.}
\label{fig:entropies}
\end{figure*}

The PDF's for the entropy rates of greedy walks are displayed in Fig.~\ref{fig:entropies}. Clearly, on average the entropy rates for greedy walks that follow the original event sequences are lower than for time-shuffled data, indicating more structured walks with repeated and more predictable node sequences. In fact, this is a direct consequence of the behavior of the coverage as a function of steps taken (Fig.~\ref{fig:coverage}) -- slower-growing coverage implies lower entropy rate of the sequence. The same applies to the fraction of back\-track\-ing steps (Fig.~\ref
{fig:bt_steps}) -- frequent back\-track\-ing steps imply high predictability and low entropy.

\begin{figure*}[ht]
\begin{center}
\includegraphics[width=0.8\linewidth]{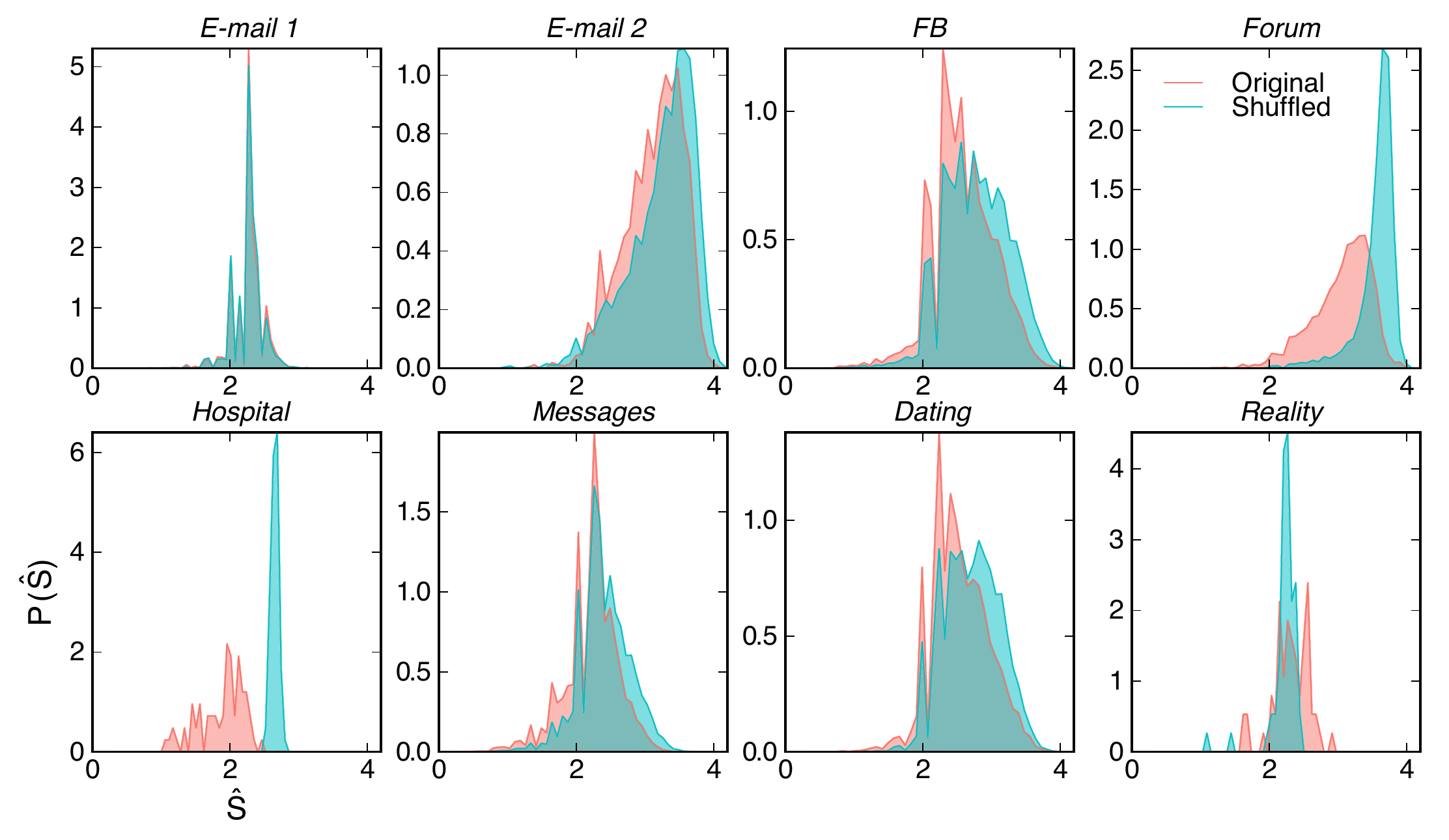}
\end{center}
\caption{Probability density functions for estimated entropy rates $\hat{S}$ of non-back\-track\-ing greedy walks for all data sets. The red lines denote PDFs for walks on top of original event sequences, whereas blue lines are again for time-shuffled reference data.
}
\label{fig:entropies_nbt}
\end{figure*}

We have repeated the same analysis for non-back\-track\-ing walks (Fig.~\ref{fig:entropies_nbt}), with a result that is in line with the coverages (Fig.~\ref{fig:coverage_nonbt}) -- entropy rates of most original data sets are still clearly below their time-shuffled counterparts (with the exception of \emph{Reality} and a vanishingly small difference for \emph{E-mail 1}). However, the differences are less pronounced than for ordinary greedy walks.

\section{Conclusions}

Studying greedy walks is a way to understand temporal networks, complementing studies of \emph{e.g.} time-respecting paths and communicability~\cite{Grindrod2011}, spreading phenomena~\cite{Karsai2011,Miritello2011}, and temporal motifs~\cite{Kovanen2011,Kovanen2013}. 
In this work, we have used two types of greedy walks, with and without backtracking, to probe the structure of time and topology in empirical temporal networks. We have shown that for all our data sets, greedy walks get trapped in non-Markovian temporal-topological structures. The clearest example is ping-pong patterns, or burst trains~\cite{Karsai2012a,KarsaiCorrelated}, of steps back and forth between two nodes. Studying the coverage statistics of non-back\-tracking walks can indicate the existence of more complex temporal-topological patterns. Also in this case, for most data sets, there are clear differences between greedy walks on real data and random null models. For example, in our data sets \emph{Forum} and \emph{Hospital}, there is a very strong suppression of the coverage for the non-back\-tracking walks. This is also reflected in an over-representation of triangle-closing steps in these data sets. In \emph{Forum}, there are triangles arising from discussions within groups of three persons (members of an Internet community); in the \emph{Hospital} data, triangles can come from two health care workers (a physician and a nurse, or two nurses) visiting a patient. By measuring the entropy rate, we put these observations on an information-theoretic basis. The conclusions from this, for our particular data sets, are the same as from the coverage statistics.

We believe that greedy walks are useful as a tool for exploring and probing temporal networks. This is not because they mimic important processes -- in practice, \emph{e.g.} spreading processes and synchronization are probably more important dynamics on temporal networks. Rather, by imposing constraints on walks such as done here, one can explore and probe temporal-topological structures in a controlled way, similarly to randomizing temporal networks in successively more restrictive ways to isolate important structures~\cite{saramaki_holme_2012}.

\section{Acknowledgments}

JS acknowledges financial support from the Academy of Finland, project n:o 260427, ``Temporal networks of human interactions''. PH was supported by Basic Science Research Program through the National Research Foundation of Korea (NRF) funded by the Ministry of Education (2013R1A1A2011947). We thank T.S. Evans for useful comments.

\bibliographystyle{epj}

\bibliography{gwalks} 

\end{document}